\documentclass[twocolumn,aps]{revtex4}
\input epsf.tex
\def\DESepsf(#1 width #2){\epsfxsize=#2 \epsfbox{#1}}
\begin{document}
\title{{\Large\bf Minimal supersymmetric SO(10) GUT with doublet Higgs}}
\author{\bf Utpal Sarkar}

\affiliation{ Deutsches Elektronen-Synchrotron DESY, Hamburg, Germany}
\affiliation{ Physical Research Laboratory, 
Ahmedabad 380 009, India}

\begin{abstract}

\noindent
We propose a simple supersymmetric SO(10) GUT model with
only doublet Higgs scalars. The fermion mass problem is
naturally solved by a new one-loop diagram.
R-parity is conserved implying a stable LSP 
which can explain the dark matter of the universe.
There are two contributions to the neutrino masses, one of
which depends on the quark mass squared while the other
is independent and similar to the type II see-saw
mechanism. In the latter case $b-\tau$ unification implies 
large neutrino mixing angles. The baryon asymmetry of the 
universe is explained through leptogenesis. 

\end{abstract}
\maketitle

The supersymmetric grand unified theories (GUTs) turned out
to be the most natural extension of the standard model, in which
all the gauge interactions unify to a single interaction. 
The simplest supersymmetric GUT
with the gauge group $SU(5)$ suffers from several problems
like proton decay, unification of coupling constants, fermion
masses and mixings. There are solutions to these problems, but
each of these solutions makes the theory more complex. The most
natural choice for the simplest GUT then turns out to 
be the one based on the group $SO(10)$. All fermions, including the
right-handed neutrinos belong to the 16-dimensional spinor
representation of the group. At higher energies the theory predicts
parity invariance and the generators of the group include
left-right symmetry and $B-L$ symmetry \cite{lr}, 
which makes the theory
even more attractive. As a consequence of the spontaneous 
breaking of the $B-L$ symmetry, the neutrino mass comes out
to be small naturally via the see-saw mechanism \cite{seesaw}.

There are many versions of the supersymmetric $SO(10)$ GUT
with varying predictions. The most popular version of the
theory includes triplet Higgs scalars to break the left-right
symmetry and simultaneously give Majorana masses to neutrinos. 
These triplets belong to a ${\bf 126}$-dimensional representation 
and a $\overline{\bf 126}$-plet
representation is required for anomaly cancellation.
Recently a minimal $SO(10)$ GUT has been proposed
\cite{min1,min2,min3,min4}, in which
the ${\bf 126}$ explains light neutrino masses and mixing \cite{min2}, 
solves the problem of wrong prediction for fermion masses in 
the $SO(10)$ GUT \cite{min1}, namely, $m_\mu = m_s$ and $m_e = m_d$,
and conserves R-parity at low energies \cite{min3} implying a stable LSP
which then solves the dark matter problem.

In the present article we propose yet another simpler 
supersymmetric $SO(10)$ GUT with only doublet Higgs
scalars. Models with doublet Higgs scalars have been discussed
in the past \cite{db,orbi}, but here we present a minimal model
with doublet Higgs scalar, which has many advantages.
The ${\bf 126}$ and $\overline{\bf 126}$ of Higgs 
scalars are now replaced by ${\bf 16}$ and $\overline{\bf 16}$ 
of Higgs scalars and one $SO(10)$ singlet fermion per
generation. This model has three more coupling
constants, but has 217 less superfields compared to the 
minimal models with triplet Higgs scalars \cite{min4}.
In addition, this model can be motivated by string theory, 
in which case the 
coupling constants will be determined by the string tension. 
This model can also be embedded in orbifold 
$SO(10)$ GUTs naturally \cite{orbi}. Leptogenesis is also possible in this
scenario, which is difficult in the minimal SUSY $SO(10)$ models
with triplet Higgs scalars. 

In the present model 
there is a new natural solution to the fermion mass problem.
Neutrino masses come out naturally light in the observed
range. Nutrino mixing angles can also be very large. In one
case the $b-\tau$ unification ensures this large mixing
angle and is consistent with small quark mixing angles.
R-parity is
also conserved in this model at low energies so that the
lightest superparticle is stable, which can then solve
the dark matter problem of the universe. 

Except for the contribution of ${\bf 126}$, 
which contains the triplet Higgs
scalars, there are many similarities between the present
model and the minimal $SO(10)$ GUT, which has been 
studied extensively during the past few years.
In the present model fermions of each generation (including a 
right-handed neutrino) belong to the 16-plet spinor representation,
which transforms under the Pati-Salam subgroup ($G_{422} \equiv 
SU(4)_c \times SU(2)_L \times SU(2)_R$) as,
$$ 
\Psi_i \equiv {\bf 16 = (4,2,1) + (\bar 4, 1, 2)} .
$$
$i = 1,2,3$ is the generation index. We also include one heavy
$SO(10)$ singlet fermion superfield $S_a \equiv {\bf 1 = (1,1,1)}$
per generation ($a=1,2,3$). When we use four numbers $(x,x,x,x)$, 
it would represent the subgroup 
$G_{3221} \equiv SU(3)_c \times SU(2)_L \times
SU(2)_R \times U(1)_{B-L}$. 

The Higgs superfields in this model are,
\begin{eqnarray}
\Phi \equiv {\bf 210} &=& {\bf (1,1,1) + (6,2,2) + (15,3,1) + (15,1,3)}
\nonumber \\ && {\bf 
+ (15,1,1) + (10,2,2) + (\overline{10},2,2)} \nonumber \\
\Gamma \equiv {\bf 16} &=& {\bf (4,2,1) + (\overline{4},1,2)} 
\nonumber \\
\overline\Gamma \equiv {\bf \overline{16}} &=& 
{\bf (\overline{4},2,1) + ({4},1,2) } \nonumber \\
H  \equiv {\bf 10} &=&{\bf (6,1,1) + (1,2,2)} . \nonumber 
\end{eqnarray}
The Higgs sector in our model is the smallest compared to all
the existing models of supersymmetric $SO(10)$ GUT. 

Before we write down the superpotential we need to discuss
the question of R-parity and matter parity in this model. 
Since the fermions contained in the superfield $\Psi$ and also the 
left-right symmetry breaking scalar superfield $\Gamma$ 
transform as ${\bf 16}$-plet superfields, we need to 
distinguish them using matter parity. Although R-parity 
and matter parity for ordinary fermions and scalars are
well defined, to include the heavier particles we need to
generalize the definition as
\begin{equation}
M = (-1)^{3(B-L) + \chi} , \label{mparity}
\end{equation}
where $\chi$ is a new quantum number and $\chi = 1$ for the
fields $S_a$, $\Gamma$ and
$\overline{\Gamma}$. Similarly, we should also extend the
definition of the R-parity and define it as,
\begin{equation}
R = (-1)^{3(B-L) + 2 S + \chi} . \label{rparity}
\end{equation}
With this extension, the R-parity now
becomes consistent with matter parity, since the fermion
superfields $S_a$ have $B-L=0$. We then need to impose that all 
interactions should be invariant under M-parity. This discrete
symmetry is not broken at any stage. 

We can then write down the superpotential with the 
scalar fields $\Phi$, $\Gamma$, $\overline{\Gamma}$
and $H$, which is invariant under M-parity as,
\begin{eqnarray}
W &=& {m_\Phi \over 4!} ~\Phi^2 
+ {m_\Gamma } ~\Gamma ~\overline{\Gamma} 
+{m_H} H^2 + {\lambda \over 4!} ~\Phi^3 \nonumber \\
&& +~ {\eta \over 4!} ~\Phi ~\Gamma ~\overline{\Gamma} 
+ H ~( \alpha ~\Gamma ~\Gamma 
+ \overline{\alpha} ~\overline{\Gamma} ~\overline{\Gamma} ) .
\end{eqnarray}
The number of parameters in this potential is same as in
the minimal SO(10) GUT with triplet Higgs scalars
\cite{min4}. 

The $SO(10)$ GUT symmetry is broken to the left-right
symmetric group $G_{3221}$ by the $vev$ of the $\Phi_1
= (1,1,1)$ and $\Phi_{15} = (15,1,1)$ components of $\Phi$
at the GUT scale $M_U$. 
The left-right symmetry is broken by the doublet Higgs scalar
$$
\xi_R \equiv (1,1,2,-1) \subset (4,1,2)
\subset \overline{\Gamma} , $$
when the neutral component of $\xi_R$ acquires a $vev$
$u_R = \langle \xi_R^0 \rangle $ at some intermediate scale.
This carries the same quantum 
number as the right-handed neutrinos. The vanishing of the
D-term is ensured by giving equal $vev$s to
$u_R$ and $\overline{u_R}= \langle \overline{\xi_R} \rangle$.
Since this field carry $B-L=-1$, this cannot give
Majorana masses to the neutrinos, but that is not a problem
as we shall discuss later. The left-handed counterpart of this
field ($\xi_L \equiv (1,2,1,-1) \subset (4,2,1)
\subset \Gamma $)
acquires a induced $vev$ after the electroweak symmetry
breaking, $u_L = \langle \xi_L^0 \rangle \sim 100$ GeV
due to the interaction $\Gamma \Gamma H$. During the left-right
symmetry breaking at a scale $u_R$, another component of
$\Phi$ acquires an induced $vev$ $\Phi_{3} = (15,1,3)$ 
due to the couplings $\Phi \Gamma \overline{\Gamma}$.

The super potential is very similar to that with triplet
Higgs and can be minimized to get the required solution in
a similar way \cite{min4}. For the symmetry breaking near the GUT scale,
we shall not include the electroweak symmetry breaking
Higgs scalar $H$. In terms of the $vev$s the superpotential
is then given by
\begin{eqnarray}
W &=& m_\Phi (\Phi_1^2 + 3 \Phi_{15}^2 + 6 \Phi_3^2)
+ 2 \lambda ( \Phi_{15}^3 + 3 \Phi_1 \Phi_{3}^2 \nonumber \\
&+& 6 \Phi_{15} \Phi_3^2) + m_\Gamma \overline{u_R } u_R
+ \eta \overline{u_R } u_R ( \Phi_1 + 3 \Phi_{15} + 6 \Phi_3) .
\nonumber
\end{eqnarray}
This equation has been solved and the possible solutions allow
the symmetry breaking scenario discussed in the previous para.
For a choice of parameters \cite{min4}, $3~\lambda ~m_\Gamma \sim
-2 ~\eta ~m_\Phi$, it allows the $SO(10)$ symmetry to break
down to the group $G_{3221}$ at the scale $M_U \sim m_\Phi$
and the symmetry breaking $G_{3221}$ to the standard model
at a scale $u_R \sim m_\Gamma$. 

All the Higgs scalars are even under matter parity and
all fermions are odd. As a result matter parity is not
broken by the $vev$ of any scalars. The R-parity for the
scalar components of the scalar superfields are always
even. So, none of the $vev$s break the R-parity. Although the
fields $\Gamma$ and $\overline{\Gamma}$ carry $B-L=-1$, 
because of the $\chi$ quantum number the R-parity is not
broken by the $vev$s of these fields. As a result at low
energy it is not possible to generate any R-parity violating
interactions. Thus the lightest superparticle will remain
stable and this can solve the problem of dark matter of the universe.

We shall now discuss the question of fermion masses.
The superpotential containing the Yukawa couplings is 
\begin{equation}
W = h_{ij} \Psi_i \Psi_j  H + y_{ia} \Psi_i S_a 
\overline{\Gamma} + M_{ab} S_a S_b + H.c.
\end{equation}
All fields are chiral superfields and hence there are no terms
with $H^*$. We imposed M-parity to write down this superpotential.
It is possible to diagonalise both $y_{ij}$ and $M_{ab}$
simultaneously. Then $y_{ij}$ contains 3 parameters, but
$h_{ij}$ contains 6 real parameters. 
Compared to the triplet Higgs models, the mass terms for the
singlets $S_a$ are the only additional parameters in the
present model. Since we diagonalised $M_{ab}$, so that 
$M_{ab} =M_{a} \delta_{ab}$ (we shall assume a hierarchy
$M_1<M_2<M_3$ and denote the scale by $M_S = M_3$), this introduces
three additional parameters compared to the minimal 
SO(10) model with triplet Higgs scalars. However,
for models with triplet Higgs, type II see-saw \cite{typeII}
contribution can work only if the model is extended 
to include another ${\bf 54}$ Higgs, which then introduces 
six more parameters \cite{goh}. In the present
model no extra field is required to get the good features
of the type II see-saw mechanism and also to generate a lepton
asymmetry of the universe.

The two 
neutral components of $H \equiv (1,2,2,0)$ acquires $vev$s
$\kappa_{u,d} \sim 10^2$ GeV, which gives masses to the
up and down sectors respectively. $\kappa_u \equiv (1,-1/2,1/2,0)$
gives masses to the up-quarks and Dirac masses to the neutrinos
while $\kappa_d \equiv (1,1/2,-1/2,0)$
gives masses to the down-quarks and charged leptons (here we 
have given the $SU(2)$ quantum numbers). Since this Higgs
does not distinguish between quarks and leptons, the quark-lepton
symmetry of the $SO(10)$ group will give a wrong mass relation
$m_\mu = m_s$ and $m_e = m_d$. In the $SO(10)$ GUTs with
triplet Higgs scalars, the $(15,2,2)$ component of ${\bf 126}$ 
receives an induced $vev$, which can solve this fermion mass
problem. In the present scenario there are no such 
representations which can solve this problem. Fortunately,
there is a one loop diagram which can solve this problem,
which we discuss below. 

\begin{figure}[h!]
\begin{center}
\epsfxsize8cm\epsffile{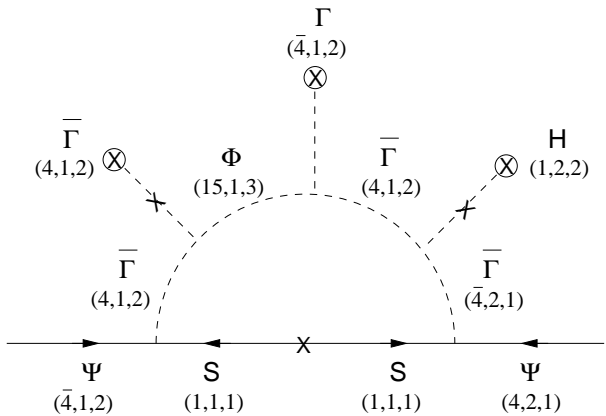}
\caption{One loop diagram contributing to the fermion masses
which breaks the quark-lepton symmetry. The components of
the fields are given below the names of the fields.}
\label{xyfig}
\end{center}
\end{figure}

Although there are no ${\bf 126}$ representation which can
contribute to the fermion masses, there is an effective term
${ y_{ia} y_{aj} \over M_{a}} ~\Psi_i ~\Psi_j 
~\overline{\Gamma} ~\overline{\Gamma} $
which comes after integrating
out the $S_a$ field. 
The $\Gamma ~\Gamma$ combination behaves like a ${\bf 126}$ 
field, and also contains a combination $(15,2,2)$, but only 
one component of this effective field gets $vev$ which 
contributes to neutrino masses only. So, this cannot solve 
the problem. Fortunately there is another effective term, which can solve
this problem,
\begin{equation}
 {\zeta \over M_X^2} ~\Psi_i ~\Psi_j 
~{\Gamma} ~\overline{\Gamma} ~H ,
\end{equation}
where $\zeta$ is some effective coupling constant and 
$M_X$ is one of the heavy scales. 
In figure 1 we presented the one
loop diagram that can contribute to the fermion masses
from this effective term. The
quantum numbers of the component fields are also presented
to demonstrate that the $(15,1,3)$ term
enters the diagram, and hence this contributes only as 
effective $(15,2,2)$ to the fermion masses. 

The contribution to the fermion masses from the one
loop diagram of figure 1 comes out to be,
\begin{equation}
\tilde M_{u,d} = \beta ~y_{ia} ~y_{aj} ~ D ~{ u_R^2 \over M_X^2} 
~ \kappa_{u,d} = D ~y_{ia}~ y_{aj} ~ m_{u,d} ,
\end{equation}
where $D = {\rm diag} [1,1,1,-3]$ is a diagonal matrix
acting on the $SU(4)_c$ space, which is the $SU(3)_c$ singlet
of ${\bf 15}$. In this estimate, the mass of the heavier
field between $S$ and $\Phi_3$ comes in the denominator. Although
most of the components of the field $\Phi$ have mass of the
order of the GUT scale, the
$(15,1,3)$ component ($\Phi_3$) remains light. The neutral
component of $(15,1,3)$ picks up an induced $vev$ 
of order $u_R$ (as discussed earlier). So, by
survival hypothesis this component remains light and its
mass will be about $m_{15} \sim (1 - 10)~ u_R$. 
So, the mass $M_X$ in the denominator of this expression
represents the heavier mass between $M_S$ and $m_{3}$.

For neutrino masses an interesting solution will correspond to
$M_S \sim u_R$, so
we take $M_X \sim 10 ~u_R$. Then including the loop factors
we get $m_{u,d} \sim 100$ MeV as required. 
The fermion masses are then given by,
\begin{eqnarray}
M_u = h \kappa_u + y^2 m_u, ~~~ &&M_d = h \kappa_d + y^2 m_d, 
\nonumber \\
M_\nu^D = h \kappa_u - 3 y^2 m_u, ~~~&& M_\ell = h \kappa_d - 3 
y^2 m_d,  \nonumber 
\end{eqnarray}
This can then solve the fermion mass problem. 

We shall now discuss the question of neutrino masses. 
Since there are no Higgs superfield with $B-L=2$, there are
no tree level Majorana neutrino masses. However, the $S_a$
singlets are Majorana particles and they couple to the
neutrinos, which can then give Majorana masses to the neutrinos.
The neutrino mass matrix in the basis $[\nu_L ~~\nu_R ~~ S]$
now becomes
\begin{equation}
M_{\nu} = \pmatrix{ 0 & h ~ \kappa_u & y ~ u_L \cr h ~ \kappa_u & 0 & y ~ u_R
\cr y~ u_L & y ~ u_R & M_a } .
\end{equation}
The entries $h, y$ and $M_a$ are $3 \times 3$ matrices (with
$M_a$ diagonal). 
The eigenvalues will now depend on the scales of $M_S$ and $u_R$.
In the limit $M_S > y~u_R$, two of them will be large with
eigenvalues
$M_S$ and $y^2~u_R^2/M_S$. The third one, which is essentially 
the left-handed neutrino, will be light and the mass is given
by,
\begin{equation}
M_{\nu ij} = { h^2 \kappa_u^2 \over v_R} 
- C h y^2 {\kappa_u^2 \over v_R} .
\end{equation}
where $v_R = y^2 u_R^2 / M_a$ and $C$ is a constant, which
determines which term dominates. 
The first term is similar to the conventional see-saw
contribution of the $SO(10)$ GUTs with triplet Higgs and proportional
to quark mass squared, while the
second contribution is similar to the type II see-saw. 

If the second term
dominates, we can follow the logic \cite{min2} similar to that of the 
triplet Higgs models to show that $b-\tau$ unification
will imply large neutrino mixing angles. From the expression
of the mass matrices we observe
$ M_\nu \propto h y^2 \propto M_d^2 - M_\ell^2 $
to a leading order. 
In the basis of diagonal charged leptons and up quarks the quark 
mixing is contained in $M_d$. Consider only second and third
generation for explaining the logic with $m_s = m_\mu = 0$. 
If the small quark mixing is
assumed to vanish, then $M_\nu \propto \pmatrix{0 & 0 \cr 0 &
m_b^2 - m^2_\tau}$. Thus with $b-\tau$ unification 
large mixing in the neutrino mass matrix will be possible
as in the type II see-saw model of the triplet Higgs $SO(10)$
GUT. 

In the limit of $M_S < y~u_R$, there are two heavy neutrinos
with masses $M_S \pm y~u_R $ and are almost degenerate. The
light neutrino will now have contribution dominantly from
the second term. 
If we consider $y~u_R \sim M_S \sim 10^{13}$ GeV and 
$h~\kappa_u  \sim y~u_L \sim 100$ GeV, then the neutrino 
masses come out to be of the order of eV. 
Including the coupling constants it will then be possible
to obtain the required neutrino masses with maximal mixings.

We now turn to the question of leptogenesis \cite{fy}. The main
constraint for thermal leptogenesis in this scenario is the
masses of the right handed neutrinos and the singlets $S_a$. 
When $M_S > y~u_R$, the right-handed neutrinos
can decay to left-handed neutrinos and Higgs bi-doublets through
lepton number violating interaction due to the effective Majorana
mass. This can then generate a lepton asymmetry of the universe.
However, since the masses of the right-handed neutrinos cannot
be smaller than $10^{13}$ GeV in this scenario or the minimal
$SO(10)$ GUT with triplet Higgs, 
the gravitino problem will not allow this mechanism to work.

In the present model this problem is solved when $M_S < y~u_R$. 
Now the decays of both the right-handed neutrinos as well as
the singlet fermions $S_a$ will contribute to leptogenesis. 
The singlets $S_a$ can decay into light leptons and Higgs
doublets $(1,2,1,-1) \subset (4,2,1) \subset \Gamma$. 
The mixing of $(1,2,1,-1)$ with ordinary Higgs bi-doublets 
$(1,2,2,0)$ after the left-right symmetry breaking will
give rise to lepton number violation. CP violation
comes from new diagrams. In this case the right-handed neutrinos
and $S_a$ will be almost degenerate.
As a result resonant leptogenesis \cite{res} will become possible.
So, the leptogenesis can take place at a much lower temperature.

The right-handed neutrinos and the $S_a$ will decay at a
higher temperature around $10^{13}$ GeV. Then inflation will
erase all asymmetry. However, after the reheating temperature
$10^{10}$ GeV, the inflaton decay will produce particles as
heavy as $10^{13}$ GeV but the number density of these particles
will be less. So, after reheating right-handed neutrinos and
$S_a$ will be produced, whose decay can generate lepton asymmetry.
Since the number density is less, there will be large suppression
in the amount of asymmetry. On the other hand, since the 
masses of the right-handed neutrinos and the singlet fermions 
are almost degenerate, there will be resonance and large
enhancement of the produced asymmetry. These two effects will
now make leptogenesis possible and it will be possible
to generate the required amount of baryon asymmetry of the 
universe before the electroweak phase transition. 

In summary, we presented a minimal supersymmetric $SO(10)$ GUT 
with only doublet Higgs scalar. This model gives correct 
symmetry breaking pattern, solves the fermion
mass problem, preserves R-parity so that the LSP can solve the
dark matter problem, predicts light Majorana neutrinos
with large mixing angle (in one case with $b-\tau$ unification) 
and able to generate a baryon asymmetry of the universe. 

I would like to thank Prof. W. Buchmuller, A. Hebecker and
K. Hamaguchi  
for discussions and acknowledge hospitality at DESY, Hamburg.
 


\end{document}